\documentclass[aps,prd,groupedaddress,nobibnote]{revtex4-1}
\usepackage{amsmath}

\DeclareMathOperator{\csch}{csch}
\usepackage{amsfonts}
\usepackage{amssymb}
\usepackage{graphicx}
\usepackage{bm}
\usepackage{hyperref}
\usepackage{caption}
\usepackage{subcaption}
\begin{document}

%\preprint{}

\title{Asymptotic States of Accelerated Qubits in Nonzero Background Temperature}


\author{A.P.C.M Lima} \email{augustopcml@gmail.com}
\author{G. Alencar}
\author{R.R. Landim}
%\thanks{}
\affiliation{Departamento de F\'{\i}sica, Universidade Federal do Cear\'{a}-
Caixa Postal 6030, Campus do Pici, 60455-760, Fortaleza, Cear\'{a}, Brazil.}

\begin{abstract}
The study of the Unruh effect naturally raises the interest for a deeper understanding of the analogy between temperature and acceleration. A recurring question is whether an accelerated frame can be distinguished from an inertial thermal bath in pure thermodynamic experiments, such problem has been approached in the literature and a consensus is yet to be fully reached. In the present work we use the open quantum system formalism to investigate the case where both acceleration and background temperature are present. We find the asymptotic state density and entanglement generation from the Markovian evolution of accelerated qubits interacting with a thermal state of the external scalar field. Our results suggest that there is a very small asymmetry on the effects of the Unruh and background temperatures. Addressing the nonzero background temperature case is of both theoretical and phenomenological interest, thus the authors hope to enrich the existing discussions on the topic.             
\end{abstract}
\maketitle

\section{Introduction}

The Minkowski vacuum, a Lorentz invariant, turns out to be seen as an planckian thermal spectrum of particles in the frame of an uniformly accelerated observer, this is the widely known Unruh Effect\cite{Unruh:1976db, Davies:1974th, Crispino:2007eb}. Being one of the key results from field quantization in general coordinates systems, it still arises in flat spacetime, without the need to include gravitational effects. The traditional treatment of the problem consists of considering  a Unruh-Dewit particle detector: a quantum particle with discrete non-relativistic energy levels weakly linearly coupled to an external quantum field. The rate of excitation of the system when travelling in an hyperbolic trajectory is then calculated and shown to be proportional to a Planck factor. The generalization of this problem for finite background temperature, which will be of importance in the present work, has been worked on in refs \cite{Padmanabhan:1988cj,Costa:1994yx}.

A more recent point of view to the examination of Unruh thermalization has been addressed by considering a similar detector and field setup in the context of open quantum systems governed by a Markovian Master Equation \cite{benati}. This approach provides a richer analysis of the evolution of the system, and the study of the phenomenon of entanglement generation for multi-particle detectors. Several generalizations to the problem have been proposed in the literature by considering different trajectories, distance between detectors, different fields, number of detectors, presence of boundaries and others (references \cite{Hu:2015lda,Zhang:2007ha,Zhang:2006ih,Yang:2016apg,Moustos:2016lol,Hu:2013ypa,Menezes:2017rby,Sokolov:2018nmu,Doukas:2010wt,She:2019hjv}). The study of entanglement is a very rich subject and associating it with Unruh Effect might provide us with further valuable  insights. Also, linking the effect to another physical quantity may become an alternative for experimental observation, which is yet to be done due to the extremely high accelerations needed to induce actually detectable thermal spectra. In ref \cite{Tian:2016gzg}, for example, it has been shown that entanglement can enhance the precision for Unruh thermalization detection. 

A subject that has caught attention recently is the study of the apparent undistinguishability between the Unruh and a thermal bath temperatures. For example,in refs \cite{Kolekar:2013hra,Kolekar:2013xua,Kolekar:2013aka} it is argued that some quantities such as density operator and detector response function are symmetric in background and Unruh temperatures, while \cite{Chowdhury:2019set} shows some aspects of distinguishability. Motivated by this problem we propose to analyze the Unruh Effect in the presence of non zero thermal background temperature in the view of open quantum systems. We keep the analysis as simple as possible, by looking at the asymptotic states generated from the system evolution and comparing the dependency of the final states on both Unruh and background temperatures we mean to provide another test on the distinguishability of the effects. It is worth mentioning that distinct features between acceleration and thermalization have already been predicted using OQS formalism, such as in the case of the presence of reflecting boundaries \cite{Zhang:2007ha} and two detectors with non vanishing distance \cite{Hu:2015lda}. Thus we focus on the simpler version(vanishing distance, no boundaries) of the setup in which the effects are originally undistinguished (when analyzed separately). 

A second motivation, shared with similar works, comes from the fact that given the general smallness of the Unruh Effect, background thermal effects are likely to be about as important as acceleration in the measures, and thus more realistic setup need to include the inertial thermal bath. Then, by extending the analysis of \cite{benati} to nonzero background temperatures, we hope to provide further insight on the study of Unruh Effect and the correlation between acceleration and temperature. 

The rest of this manuscript will be divided in 3 sections. In the next we present a review of the formalism from \cite{benati}, which we use as basis. In section 3 we expose our results for the case of finite temperatures. Section 4 will consist of our final remarks on the problem.
\section{Markovian Evolution of detectors in Minkowski Vacuum}

We begin with a brief review of the method and results from \cite{benati}, which we use as the start point for our analysis. The system in consideration is a two-level Unruh-Dewitt detector weakly coupled to an external massless scalar field. Instead of the standard transition rate calculation, the state density operator of the entire system (detector+field) is considered, and a reduced dynamics for the detector's state density is derived.

\subsection{Time Evolution for a Single Particle}
The total system Hamiltonian is:
\begin{equation}
H_s=H_D+H_f+H_I,
\end{equation}
where $H_d$ and $H_I$ are respectively, the detector,field and interaction Hamiltonians given by
\begin{align}
H_D=\frac{\omega}{2}\sigma_z\otimes \mathbf{1},\nonumber\\
H_I=\lambda \sigma_\mu \otimes \Phi_\mu,
\end{align}
with $H_f$ being the usual free scalar field Hamiltonian. The field $\Phi_\mu$ is defined as
\begin{equation}
\Phi_\mu=\chi_\mu\phi^-+\chi_\mu^*\phi^+,
\end{equation}
where $\chi_\mu$ serve as generalized coupling constants, $\sigma_0$ is the identity $2x2$ and $\sigma_i$ the Pauli matrices, $\phi^{\pm}$ are the positive and negative parts of the field expansion. For simplicity, these constants will be assumed to satisfy
\begin{equation}
\chi_\mu\chi_\nu=\delta_{\mu\nu}.
\end{equation}
The initial density operator of the system will be taken to be
\begin{equation}
\rho_{Total} (0)=\rho(0)\otimes \rho_F(0),
\end{equation}
where $\rho$ and $\rho_F$ represent the densities of the detector and the field, respectively. 

A properly defined tracing procedure over the field degrees of freedom leads to an expression for the time evolution of the detector's state density, however, it is a very complicated equation containing memory effects. By requiring that $\rho_F$ is a static configuration of the field over the time evolution generated by $H_F$ and the condition of weak coupling, formally carried out at \cite{davies,davies2}, one obtains a Markovian (no memory effects) equation of motion, known as the Kossakowski-Lindblad equation:
\begin{equation}\label{master}
\dot{\rho}=-i[H_{eff},\rho(t)]+\mathcal{L}[\rho(t)].
\end{equation}  
The operator $\mathcal{L[\rho]}$ in the above equation is the Lindbladian, given by
\begin{subequations}\label{lind}
\begin{align}
&\mathcal{L}[\rho]=\frac{1}{2}\sum_{i,j=1}^3 a_{ij}[2\sigma_j\rho\sigma_i-\sigma_i\sigma_j\rho-\rho\sigma_i\sigma_j],\\
&a_{ij}=A \delta_{ij}-iB\epsilon_{ijk}+Cn_in_j.
\end{align}
\end{subequations}
The coefficients $A,B$ and $C$ in the Kossakowski-Lindblad matrix $a_{ij}$ are given in terms of Fourier transforms of the field Wightman functions
\begin{subequations}
\begin{align}
&\mathcal{G}(\omega)=\int_{-\infty}^{\infty}dt\exp[i\omega t]\langle\phi[x(t)]\phi[x(0)]\rangle,\label{trans}\\
&A=\frac{1}{2}[\mathcal{G}(\omega)+\mathcal{G}(-\omega)],\\
&B=\frac{1}{2}[\mathcal{G}(\omega)+\mathcal{G}(-\omega)],\\
&C=\frac{1}{2}[\mathcal{G}(0)-\mathcal{G}(\omega)-\mathcal{G}(-\omega)], 
\end{align}
\end{subequations}
and
\begin{subequations}
\begin{align}
&H_{eff}=\frac{\Omega}{2}\vec{n}\cdot\vec{\sigma},\\
&\Omega=\omega+i[\mathcal{K}(-\omega)-\mathcal{K}(\omega)],\label{effec}\\
&\mathcal{K}(\omega)=\frac{1}{i\pi}P\int_{-\infty}^\infty d\lambda \frac{\mathcal{G}(\lambda)}{\lambda-\omega},\label{lambs}
\end{align}
\end{subequations}
with $P$ denoting principal value.

When $\rho_F(0)$ is Minkowski's vaccum, we have: 
\begin{equation}\label{wight}
\langle\phi(x+\Delta x)\phi(x)\rangle=G(x+\Delta x,x)=-\frac{1}{4\pi^2}\frac{1}{(\Delta x^0-i\epsilon)^2-\vec{\Delta x}^2}.
\end{equation} 
The detector is moving on hyperbolic trajectory with proper time $t$:
\begin{subequations}\label{trajec}
\begin{align}
&x^0(t)=\frac{1}{a}\sinh(at),\\
&x^1(t)=\frac{1}{a}\cosh(at),\\
&x^2=x^3=0.
\end{align}
\end{subequations}
Inserting this trajectory in (\ref{wight}) and integrating in (\ref{trans}) we obtain:
\begin{subequations}
\begin{align}
&\mathcal{G}(\omega)=\frac{1}{2\pi}\frac{\omega}{1-e^{-\beta_U\omega}},\\
&A=\frac{\omega}{4\pi}\left(\frac{1+e^{-\beta_U\omega}}{1-e^{-\beta_U\omega}}\right),\\
&B=\frac{\omega}{4\pi},\\
&C=\frac{\omega}{4\pi}\left(\frac{2}{\beta_U\omega}-\frac{1+e^{-\beta_U\omega}}{1-e^{-\beta_U\omega}}\right).
\end{align}
\end{subequations}
By pluging these back in (\ref{lind}) and conveniently decomposing the state density of the detector in the Bloch vector
\begin{equation}
\rho=\frac{1}{2}\left(1+\sum_i\rho_i\sigma_i\right),
\end{equation}
one obtains the time evolution
\begin{equation}\label{evo1p}
\frac{\partial}{\partial t}\vert\rho(t)\rangle=-2\mathcal{H}\vert\rho(t)\rangle+\vert \eta\rangle
\end{equation}
where $\eta_i=-4 B n_i$ and
\begin{equation}
\mathcal{H}=\left(
\begin{array}{ccc}
a&b+\Omega_3&c-\Omega_2\\
b-\Omega_3&\alpha&\beta+\Omega_1\\
c+\Omega_2&\beta-\Omega_1&\gamma
\end{array}
\right),
\end{equation}
where
\begin{align}
&a=2A+C(n_2^2+n_3^2), \quad b=-Cn_1n_2,\nonumber\\
&\alpha=2A+C(n_1^2+n_3^2), \quad c=-Cn_1n_3,\nonumber\\
&\gamma=2A+C(n_1^2+n_2^2), \quad \beta=-Cn_2n_3,\nonumber\\
&\Omega_i=\frac{\Omega}{2}n_i,\quad n=1,2,3.
\end{align}
The solution of eq (\ref{evo1p}) allows one to compute, for example, the usual response function for the 2-level Unruh-Dewitt detector, here for simplicity and conciseness we will focus on the assymptotic state($t\rightarrow\infty$):
\begin{equation}\label{assy}
\rho_i=-\frac{B}{A}n_i=\tanh\left(\frac{\beta_U\omega}{2}\right)n_i.
\end{equation} 
This state density can be identified with the one for thermal equilibrium in a $1/\beta_U$ temperature heat bath:
\begin{equation}
\rho_{\beta}=\frac{e^{-i\beta_U H_D}}{Tr[e^{-i\beta_U H_D}]}.
\end{equation}
\subsection{Two Particles and Entanglement Generation}
For the case of two particles following the same trajectory (\ref{trajec}), the system will also be described by an equation of the type (\ref{master}), with the Lindbladian given by
\begin{align}
\mathcal{L}[\rho]=&\sum_{i,j}a_{ij}\left(\left[(\sigma_j\otimes\sigma_0)\rho(\sigma_i\otimes\sigma_0)-\frac{1}{2}\{\sigma_i\sigma_j\otimes\sigma_0,\rho\}\right]\right.\nonumber\\
&+\left.\left[(\sigma_0\otimes\sigma_j)\rho(\sigma_0\otimes\sigma_i)-\frac{1}{2}\{\sigma_0\otimes\sigma_i\sigma_j,\rho\}\right]\right.\nonumber\\
&+\left.\left[(\sigma_j\otimes\sigma_0)\rho(\sigma_0\otimes\sigma_i)-\frac{1}{2}\{\sigma_i\otimes\sigma_j,\rho\}\right]\right.\nonumber\\
&+\left.\left[(\sigma_0\otimes\sigma_j)\rho(\sigma_i\otimes\sigma_0)-\frac{1}{2}\{\sigma_j\otimes\sigma_i,\rho\}\right]\right).
\end{align}
The $a_{ij}$ coefficients are the same as in the single particle case and the new effective Hamiltonian is given by: 
\begin{equation}
H_{eff}=H_{eff}^{(1)}+H_{eff}^{(2)}+H_{eff}^{(1,2)},
\end{equation}
where $H_{eff}^{(1)}$ and $H_{eff}^{(2)}$ are defined analogously to (\ref{effec}) and 
\begin{equation}
H_{eff}^{(1,2)}=i\sum_{ij}\{[\kappa(\omega)+\kappa(-\omega)]\delta_{ij}+[\kappa(0)-\kappa(\omega)-\kappa(-\omega)]n_in_j\}.
\end{equation} 
The term above represents an indirect interaction, however, from the definition (\ref{lambs}) we see that $\kappa(\omega)$ is an odd function and thus the term cancels out, leaving just the individual and acceleration independent Lamb-shifts. For further simplification and analysis of the acceleration induced effects, the effective Hamiltonian will be dropped from (\ref{master}) in this case. 

Decomposing the reduced density as
\begin{equation}
\rho(t)=\frac{1}{4}[\sigma_0\otimes\sigma_0+\sum_i\rho_{0i}(t)\sigma_0\otimes\sigma_i+\sum_i \rho_{i0}(t)\sigma_i\otimes\sigma_0+\sum_{i,j}\rho_{ij}(t)\sigma_i\otimes\sigma_j],
\end{equation} 
the following equations are obtained: 
\begin{subequations}
\begin{align}
&\dot{\rho}_{0i}=-4A\rho_{0i}+2B(2+T)n_i-2B\sum_kn_k\rho_{ik},\\
&\dot{\rho}_{i0}=-4A\rho_{i0}+2B(2+T)n_i-2B\sum_kn_k\rho_{ki},\\
&\dot{\rho}_{ij}=-4A[2\rho_{ij}+\rho_{ji}-T\delta_{ij}] ,\nonumber\\
&+2B[n_i(2\rho_{0j}+\rho_{j0})+n_j(2\rho_{i0}+\rho_{0i})-\delta_{ij}\sum_kn_k(\rho(\rho_{k0}+\rho_{0k})], 
\end{align}
\end{subequations}
which in the assymptotic state reduce to 
\begin{subequations}\label{assymp}
\begin{align}
\hat{\rho}_{i0}&=\hat{\rho}_{0i}=\frac{R(T+3)}{3+R^2}n_i,\\
\hat{\rho}_{ij}=\hat{\rho}_{ji}&=\frac{1}{3+R^2}[n_in_j(T+3)+(T-R^2)\delta_{ij}].
\end{align}
\end{subequations}

To analyze the entangling of the two qubits, one can compute the Concurrence\cite{Hill:1997pfa,Wootters:1997id,concurrence}. Let $\rho_c$ be defined by
\begin{equation}
\rho_C=\rho(\sigma_2\otimes\sigma_2)\rho^{T}(\sigma_2\otimes\sigma_2),
\end{equation}
with positive eigenvalues $\lambda_{1,2,3,4}$, the concurrence $\mathcal{C}$ is then given by
\begin{equation}
\mathcal{C}[\rho]=max\{\lambda_1-\lambda_2-\lambda_3-\lambda_4,0\},
\end{equation}
Assuming values $0\leq \mathcal{C}\leq 1$. With unity value representing a completely entangled state and zero for a completely untangled one. 

The concurrence for the asymptotic configuration in (\ref{assymp}) is given by
\begin{equation}\label{conc}
\mathcal{C}[\hat{\rho}]=\frac{3-R^2}{2(3+R^2)}\left[\frac{5R^2-3}{3-R^2}-T\right],
\end{equation}  
for $T<(5R^2-3)/(3-R^2)$, or zero otherwise.

As a simple example, one can consider the two detectors prepared in an originally separable state($\mathcal{C}=0$): 
\begin{equation}
\rho(0)=(1+\vec{p}\cdot\vec{\sigma})\otimes(1+\vec{q}\cdot\vec{\sigma}),
\end{equation}
with $\vec{q}$ and $\vec{p}$ denoting unit vectors. Then, the maximum entanglement in the final state will be achieved for $T=\vec{q}\cdot\vec{p}=-1$ and will be given by
\begin{equation}\label{ema}
\mathcal{C}[\hat{\rho}]=\frac{2R^2}{3+R^2},
\end{equation} 
illustrating thus, the entanglement generation and it's dependency with the acceleration. The results reviewed in this section are identical to those one obtains from considering static particles in a common thermal bath.

Next, we want to analyze the case where besides an acceleration, there is also a finite background temperature. Our main focus will be the question as to whether the effects induced by acceleration and temperature are distinguishable, and if so, how different these effects are. 
 
\section{Finite Temperature Background}

It has been shown in earlier works that under certain circumstances, uniform acceleration will have distinct features from a inertial thermal bath in the state density evolution of the detectors, such as the presence of reflecting boundaries and finite distance between detectors. Here, instead we consider the simpler case reviewed in the last section, and introduce a background uniform temperature as measured by an inertial observer. We want to compare, in the case where both effects are present, the temperature and acceleration dependencies of the process, as well as having a look at the summed effect.

The initial configuration of the field will be given by
\begin{equation}
\rho_f(0)=\frac{e^{-\beta H_f}}{Tr[e^{-\beta H_f}]},
\end{equation}
which is an equilibrium state of the field's natural evolution $[\rho_f(0), H_f]=0$. In this configuration, the correlation function for the field is
\begin{equation}
G^+(x+\Delta x,x)=-\frac{1}{4\beta}\sum_{n=-\infty}^\infty\frac{1}{(\Delta x^0-i\epsilon)^2-\vec{\Delta x}^2+i n \beta}.
\end{equation} 
For a static trajectory it reduces to the same expression as in the accelerated case with $2\pi\beta\rightarrow a$. The explicit form for a trajectory of type (\ref{trajec}) is worked out in \cite{Costa:1994yx}(se also the explicit form in \cite{Weldon:2000pe}), for the case $\Delta x=x(t)-x(0)$ it reduces to
\begin{equation}\label{wight2}
G^+(t,0)=\frac{1}{8\beta\alpha}\csch^2\left(\frac{\pi t}{\alpha}\right)\left\{\coth\left[\frac{\alpha}{\beta}(e^{2\pi t/\alpha}-1)\right]-\coth\left[\frac{\alpha}{\beta}(1-e^{-2\pi t/\alpha})\right]\right\},
\end{equation}
where $\alpha$ and $\beta$ are the inverse of the unruh and background temperatures respectively and the $i\epsilon$ factors have been suppressed. 

\begin{figure}
\centering
\includegraphics[width=10cm]{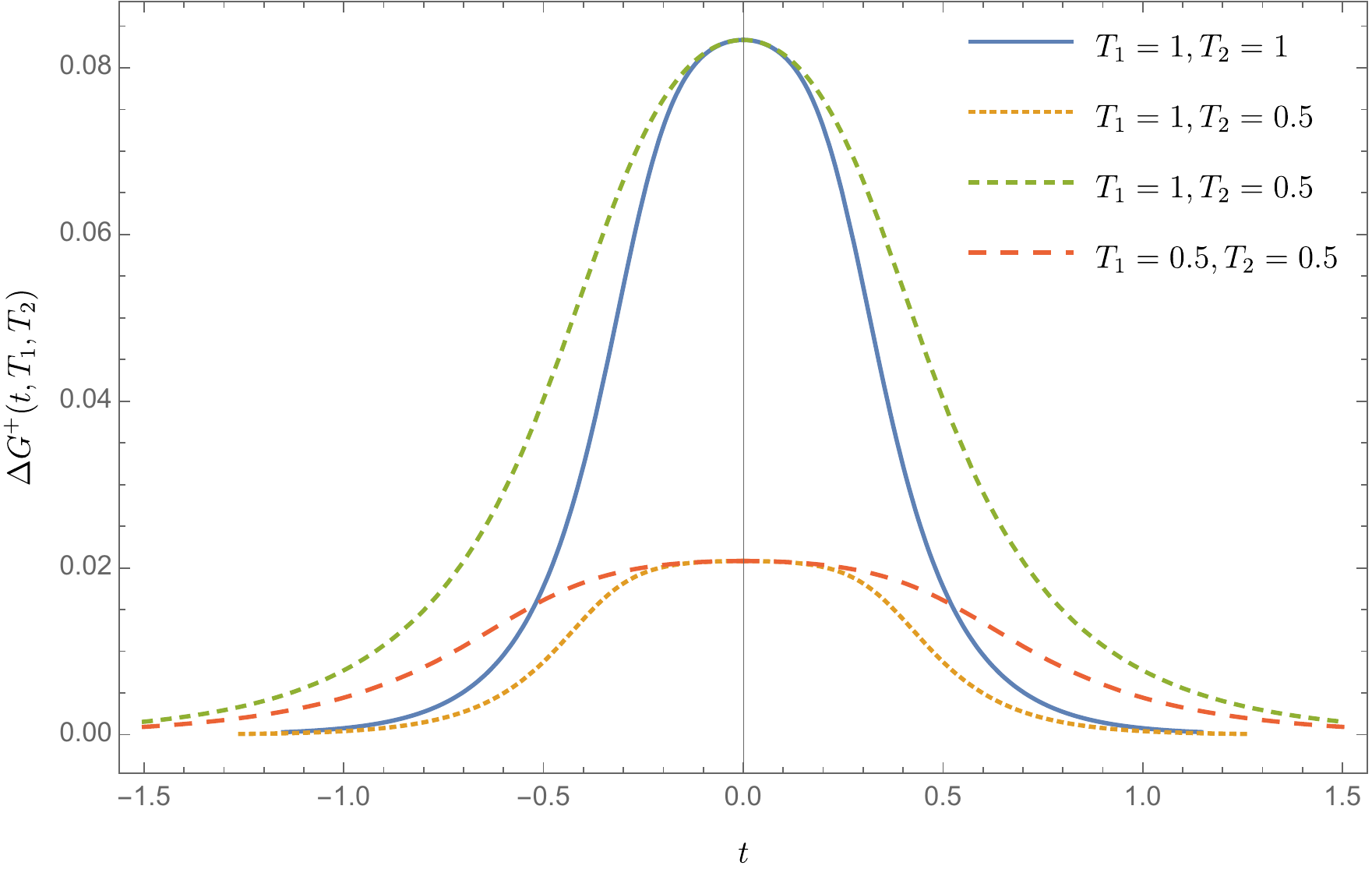}
\caption{ Plots of $\Delta G^+_\beta(t,T_1,T_2)$. The curves slightly resemble a Gaussian, with peak value $T_1^2/12$ at $t=0$. Notice that for same $T_1$, the curve characteristic width will be decreasing with $T_1$.}
\label{fig1}
\end{figure}

Unfortunately the Fourier transform for expression (\ref{wight2}) doesn't seen to have an closed expression. We employ then a trick from \cite{Costa:1994yx}, by separating a background temperature "correction":
\begin{equation}\label{deco}
G^+(t,0)=-\frac{1}{4\alpha^2}\csch^2(\pi t/\alpha)+\Delta G_{\alpha\beta}^+,
\end{equation} 
where
\begin{equation}\label{deco2}
\Delta G_{\alpha\beta}^+=\frac{1}{4\alpha^2}\csch^2(\pi t/\alpha)+G^+.
\end{equation}
This simply factors out the pure acceleration part which we already know how to treat. The remaining contribution $\Delta G_{\alpha\beta}^+$ is well behaved( on the sense that it's singularity at $t=0$ is removable) on the real axis, and thus we can drop the $i\epsilon$ factors and easily integrate it numerically(see the plots in figure \ref{fig1}). Equivalently one could instead separate the pure temperature term and deal with the remaining part(see (\ref{deco3})). 

From the expressions for the asymptotic state densities (\ref{assy},\ref{assymp}) we can narrow the analysis to the examination of the quantity $R=B/A$. Conveniently, $G_\beta^+$ is an even function of $t$, so that $B$ remains unchanged, for $A$ we have
\begin{equation}
A\rightarrow A+\int_{-\infty}^\infty {dt} \cos(\omega t)\Delta G_{\alpha\beta}^+.
\end{equation}
Then, let the quantity $\gamma(\omega,\alpha,\beta)$ be defined by
\begin{equation}
\gamma(\omega,\alpha,\beta)=\int_{-\infty}^\infty {dt} \cos(\omega t)\Delta G_{\alpha\beta}^+,
\end{equation}
so $R$ can be writen as
\begin{equation}\label{R}
R(\omega,\beta^{-1},\alpha^{-1})=\left[\coth\left(\frac{\alpha\omega}{2}\right)+\frac{4\pi}{\omega}\gamma(\omega,\alpha,\beta)\right]^{-1}.
\end{equation}
From equation (\ref{wight2}) one would guess acceleration and background temperature have very distinguishable effects, however, as we will see shortly the asymmetry of $R$ in the two temperatures is generally very small.

\begin{figure}[ht]
  \centering
  \begin{subfigure}[h]{0.5\linewidth}
    \centering\includegraphics[width=8cm]{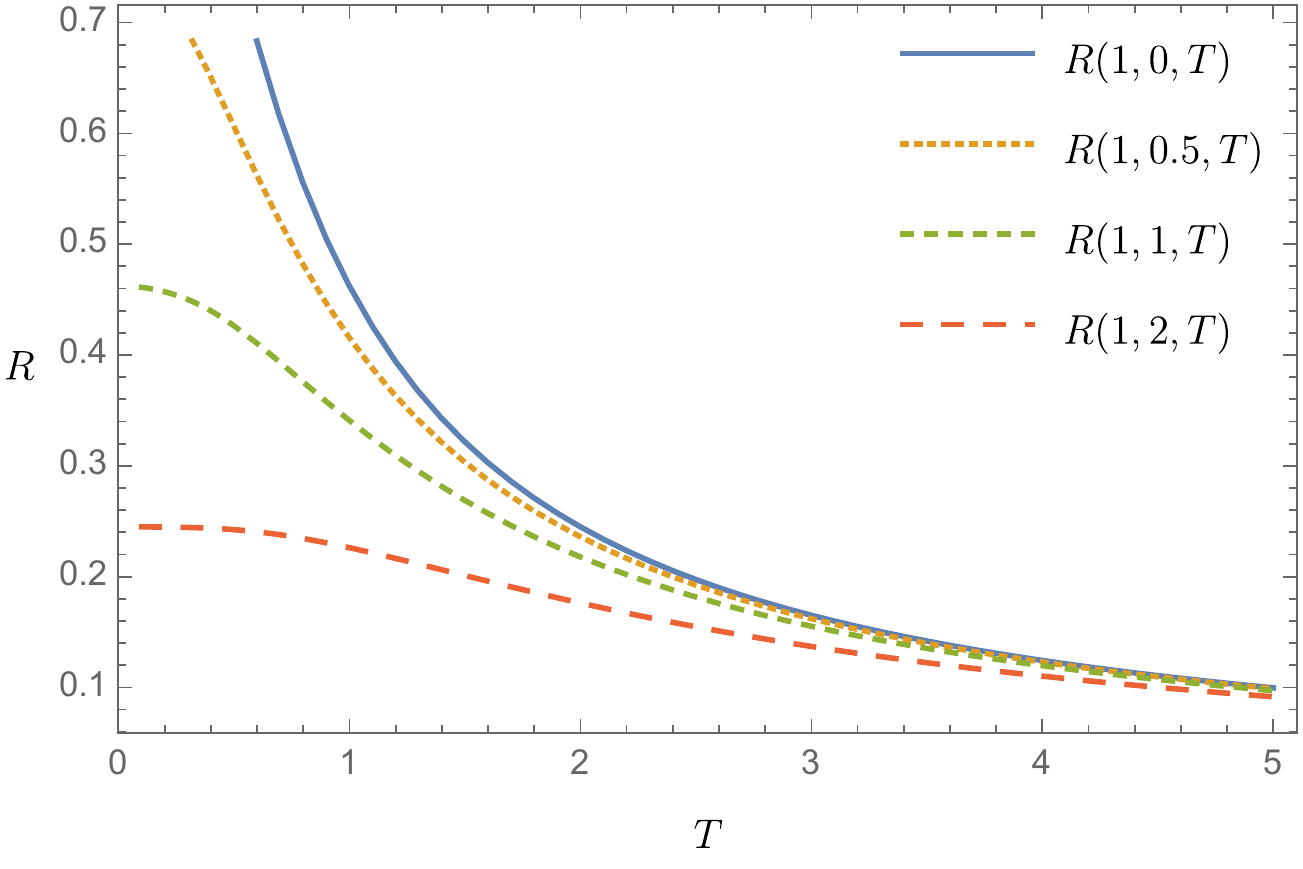}
    \caption{Plot of $R(1,T',T)$ against $T$ for a few values of $T'$.\label{fig2}}
  \end{subfigure}%
  \begin{subfigure}[h]{0.5\linewidth}
    \centering\includegraphics[width=8cm]{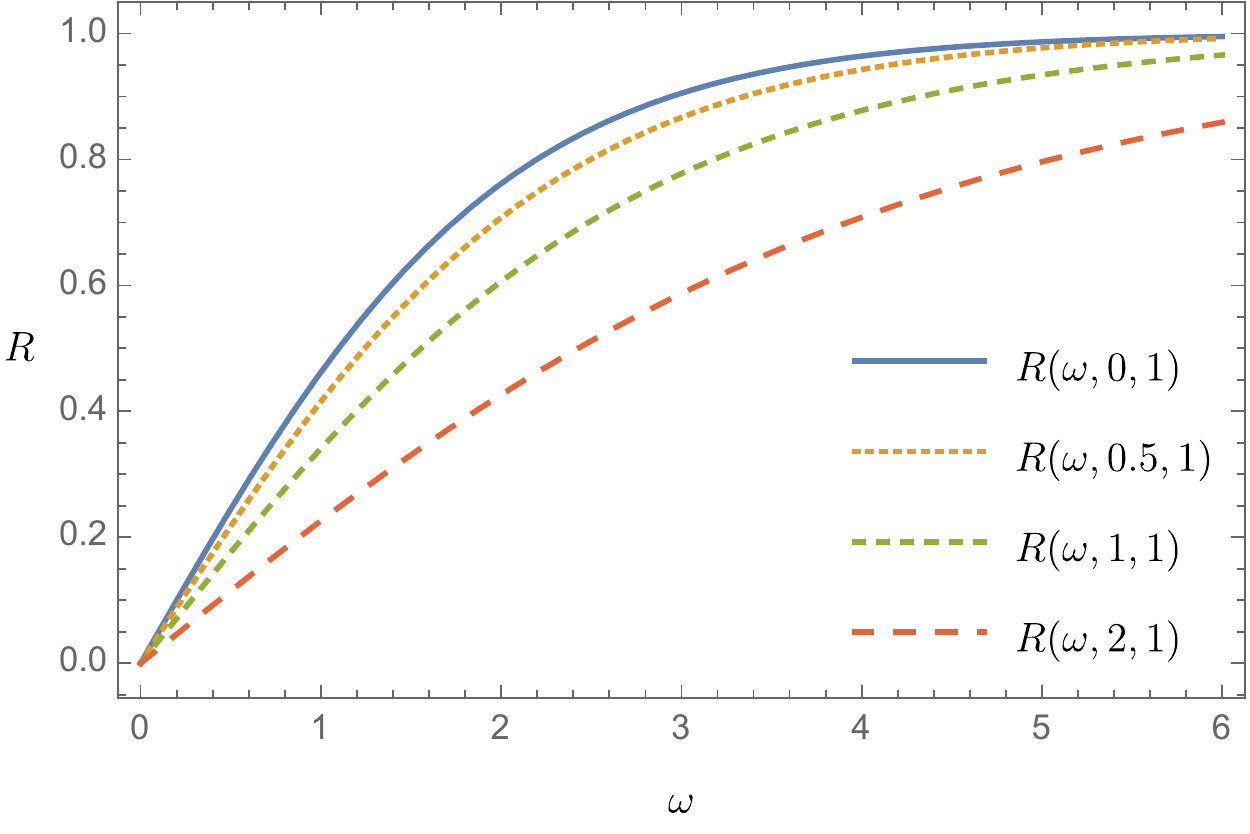}
    \caption{Plot of $R(\omega,T,T')$ against $\omega$ for a few values of $T'$.\label{fig3}}
  \end{subfigure}
  \caption{Plotting $R(\omega,T_1,T_2)$ against temperature and detector frequency.}
  \label{fig23}
\end{figure}

Our main focus will be the analysis of general features, so we will be using simplistic values on the arguments of (\ref{R}) rather than more experimentally realistic ones(see \ref{real}). First lets have a look at the acceleration dependency of $R$ for a few different background temperatures(figure \ref{fig2}). It can be seen that the values decrease progressively with background temperature for small accelerations.

\begin{figure}[ht]
  \centering
  \begin{subfigure}[h]{0.5\linewidth}
    \centering\includegraphics[width=8cm]{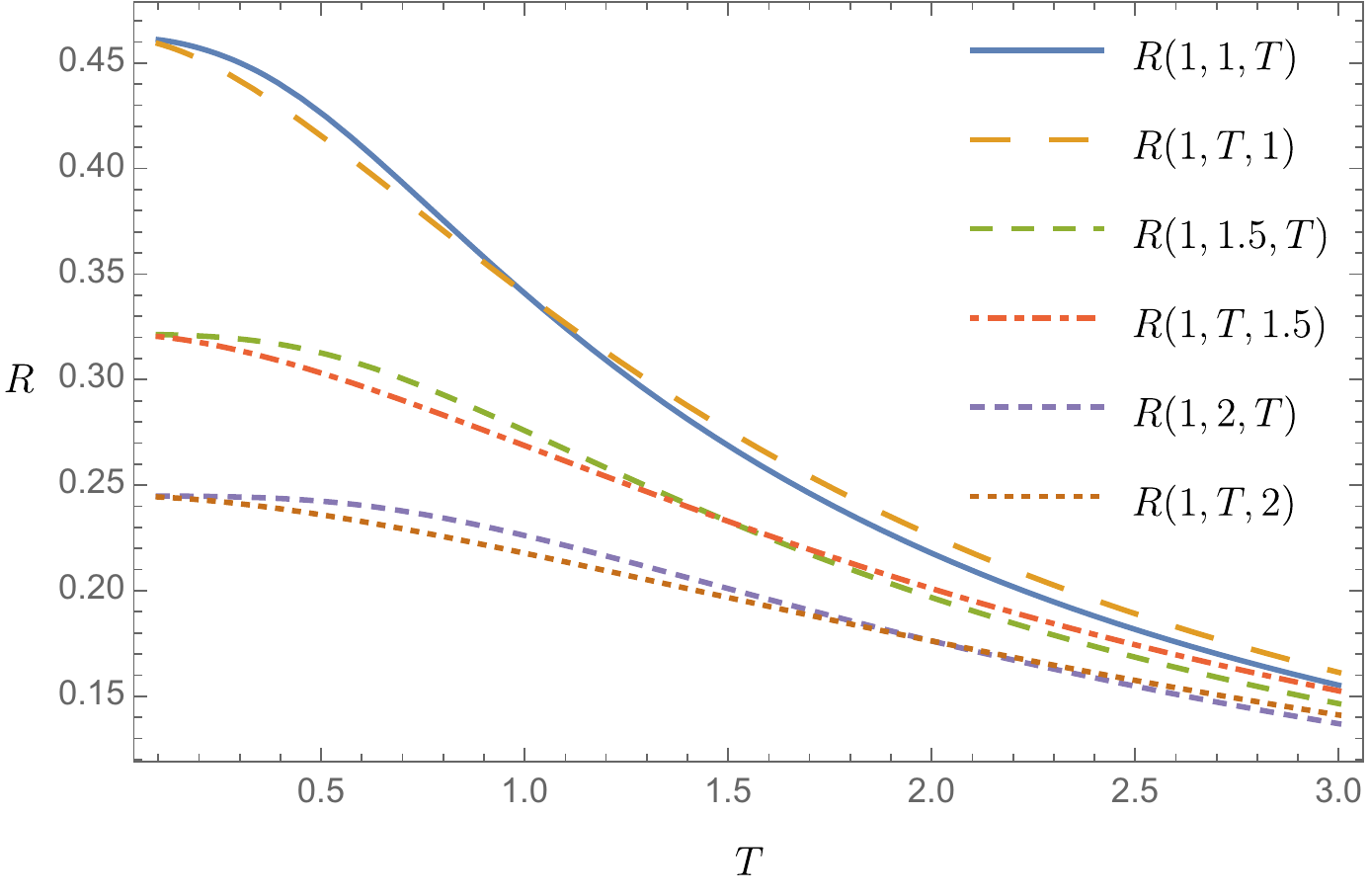}
    \caption{Values of $R(1,T,T')$ and $R(1,T',T)$ for a few examples of $(T,T')$.\label{fig4}}
  \end{subfigure}%
  \begin{subfigure}[h]{0.5\linewidth}
    \centering\includegraphics[width=8.5cm]{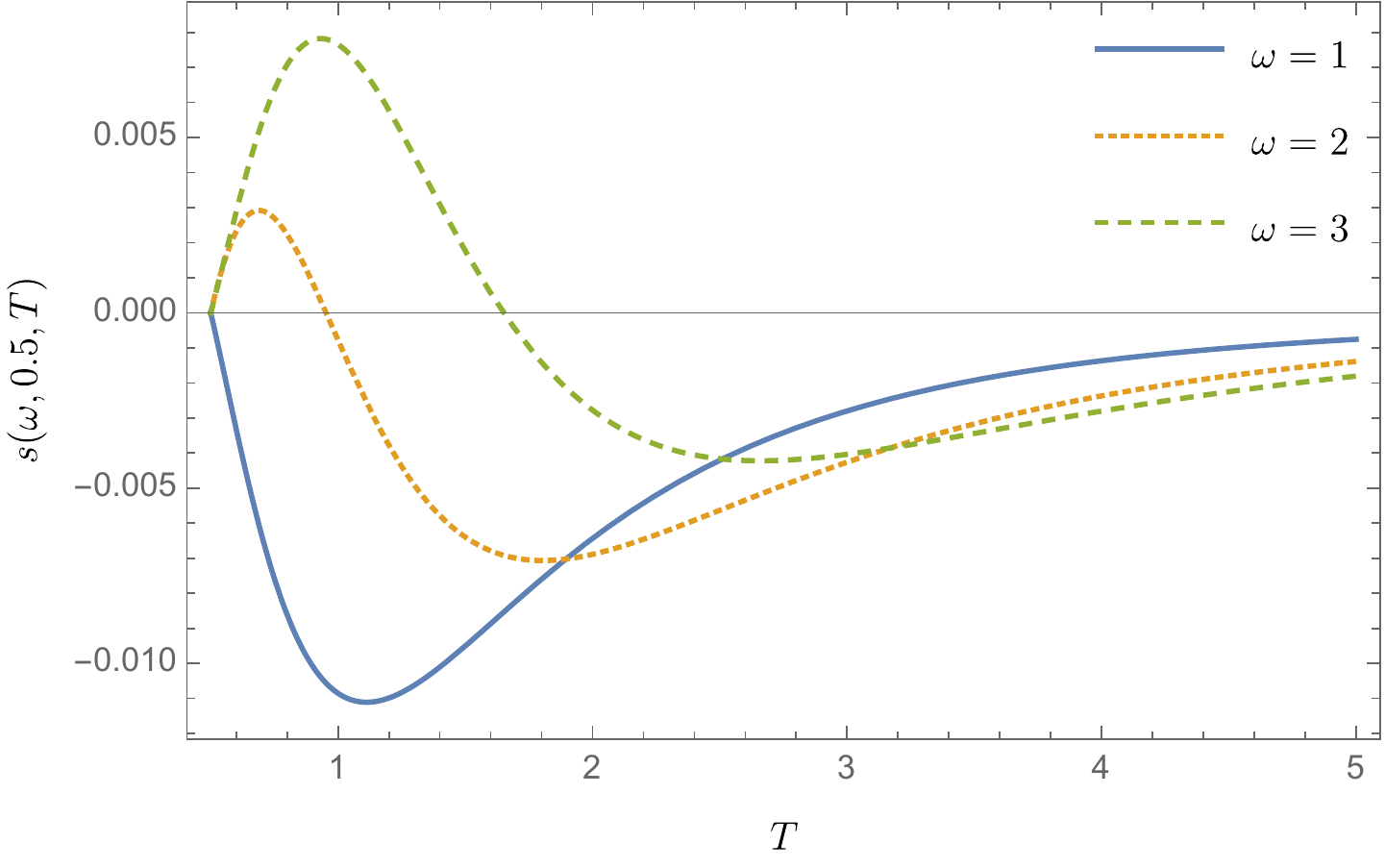}
    \caption{Values of $s(\omega,0.5,T)=R(\omega,0.5,T)-R(\omega,T,0.5)$ for $\omega=1,2,3$.\label{fig5}}
  \end{subfigure}
  \caption{Asymmetry of $R$ in the Unruh and background temperatures.}
  \label{fig45}
\end{figure}

Next we plot the values of $R$ for different detector frequency values. The curves in (figure \ref{fig3}) suggest an monotonical dependency, for smaller values of $\omega$ we have an approximately linear growth. Now to check what happens when we swap the background and Unruh temperatures, the comparative curves are plotted in (figure \ref{fig4}). As mentioned before, just by looking at the Green's function (\ref{wight2}) one could be naively led to think there would be a big difference on the values of R obtained by swaping $\alpha$ for $\beta$. In (figure \ref{fig5}) we can observe the values of $s(\omega,T_1,T_2)=R(\omega,T_1,T_2)-R(\omega,T_2,T_1)$. 

Notice that the equivalent acceleration/temperature curves have very little deviations from one another. By analyzing the temperature and frequency dependencies, we can infer that this ``closeness'' feature is quite general. First suppose we tinker with the frequency parameter, notice that for high frequency values, $R$ asymptotically approaches unit value, and for low frequencies the values vary almost linearly(figure \ref{fig3}), thus generating effects almost proportional to those in figure \ref{fig4}. Now we manipulate temperatures, if Unruh and background temperature are too different, say $T>>T'$, then $R(\omega,T',T)$ will be close to $R(\omega,0,T)$ and thus there won't be a much bigger distinguishability either. Last, changing both temperatures proportionally can be equivalently done by switching time scale and changing frequency instead, so it falls back to the $\omega$ analysis(see \ref{real}).

%\begin{figure}[h]
%\center
%\subfigure[fig2][Plot of $R(1,T',T)$ against $T$ for a few values of $T'$]{\includegraphics[width=8cm]{fig2.pdf}}
%\qquad
%\subfigure[fig3][Plot of $R(\omega,T,T')$ against $\omega$ for a few values of $T'$]{\includegraphics[width=8cm]{fig3.pdf}}
%\label{fig23}
%\caption{Plotting $R(\omega,T_1,T_2)$ against temperature and detector frequency.}

%\end{figure}

%\begin{figure}[h]
%\label{fig45}
%\center
%\subfigure[fig4][Values of $R(1,T,T')$ and $R(1,T',T)$ for a few examples of $(T,T')$]{\includegraphics[width=8cm]{fig5.pdf}}
%\qquad
%\subfigure[fig5][Values of $s(1,T,0.5)=R(1,T,0.5)-R(\omega,0.5,T)$]{\includegraphics[width=8cm]{fig4.pdf}}
%\caption{Asymmetry of $R$ in the Unruh and background temperatures.}

%\end{figure}

\subsection{Discussing realistic values}\label{real}

Up to now we have only used rather fictitious values for the parameters $\omega,\alpha$ and $\beta$, ignoring the magnitude order at which these quantities might seen plausible to consider. This approach has been used to facilitate our analysis by generating more sensible results.

To get a look at more realistic values we need restore the constants involved. First, the frequency values, for a two-level system with energy  gap $E$, we will have $\omega=E/\hbar$. Thus for a frequency of say, $1 eV$, the frequency time argument will be of $10^{15}$ order. It is known that very high accelerations are needed to generate even a seemingly small Unruh temperature, however, let's assume the acceleration induced thermalization stands on the same, or close to the order of a plausible background temperature as to generate a measurable effect. The general time coefficient $k_B T/\hbar$ in the arguments of the Green's function will be of order $10^{11} \times O[T]$. 

From the ``thermal correction'' part of the Green's function (\ref{deco2}) we can see that the detector only effectively interacts with the background temperature for a short time period( this is not really accurate as this interaction is not factored out, but forcefully separated in the expression), the width of the interval is proportional to the acceleration. Also, the usual part of Unruh thermalization has the argument proportional to $\omega/a$ (or $\omega/T$). Then, we are able to simplify the arguments in (\ref{R}) by adjusting the units of the usual time-scales. In other words, we have $R(\omega,T,T')=R(\lambda\omega,\lambda T,\lambda T')$ for any positive $\lambda$, as can be easily seen from (\ref{deco2},\ref{R}). The close to unity values used earlier correspond to $\alpha\sim\beta$ and $\beta\omega\sim k_B$, this is achievable for a temperature of $1K$ using energy gaps on the hydrogen fine structure order. Unfortunately the acceleration values corresponding to this temperature are still too high, and a direct observation would require a much more sensible(low energy) detector. A more realistic option would be to consider circular trajectories(see \cite{Costa:1994yx,She:2019hjv}), which could be realized in the future using more potent particle accelerators.      

Usually, it would be reasonable to expect background temperature values to be higher than the acceleration induced effect. In this case, it would be most efficient to separate the field Green's function as
\begin{equation}\label{deco3}
G^+=-\frac{1}{4\beta^2}\csch^2(\pi t/\beta)+\Delta G_{\beta\alpha}^+,
\end{equation}  
where $\Delta G_{\beta\alpha}^+$ is defined in analogy to (\ref{deco2}). This way, we generate a small acceleration ``correction'' to the inertial thermalization. This is advantageous because in the (\ref{deco}) case, the width of the usual time scale is increasing in $\alpha$, whereas in the above case it increases with $\beta$, making it easier to do precise integrations by picking whichever factor has a higher value. 

\section{Final Remarks}
 
 In the present work we proposed an extension to the problem presented in \cite{benati} by inserting a nonzero background temperature on the analysis of Unruh-Dewit detectors as open quantum systems. We proceeded by keeping the calculations as simple as possible, looking at the asymptotic states obtained from a Markovian evolution of the particle detectors. As we did not find a closed expression for the Fourier transform of the external field correlations, the results are obtained numerically (figures \ref{fig23} and \ref{fig45}). Also, we used these results as a further test to the acceleration-temperature correspondence.
  
 The presence of a background temperature, as expected, contributed to further decrease entanglement generation between detectors, and it's temperature/acceleration evolution curves look pretty familiar to the usual Unruh Effect. On testing whether the asymptotic states were symmetrical in exchanging background and Unruh temperatures, we found that there is a small difference in the plotted curves, further suggesting that a common thermal bath and the acceleration induced one have distinct, although very close, properties. Since the analysis provided is pure Markovian, including memory effects and finite time behaviors might be of key importance and are to be approached in future works. Besides the close analogy between Unruh effect and thermal baths, the authors do not know why the corresponding curves found in figure \ref{fig4} are so proximate to each other. 
 
The acceleration needed for the Unruh-Effect to be actually measurable is unfornately too high to achieve with current tecnology. A reasonable work around proposed would be to accelerate the particle detector in a circular trajectory (an open quantum system approach to Unruh effect in cicular trajectory can be found in \cite{She:2019hjv}). But even in very high acceleration scenarios, one would expect background temperature to be of considerable order and thus it has to be considered on further works that propose to tackle an realistic view of the problem. 
 
 Although the analysis provided is very simplified and the parameters used are rather unrealistic, the authors hope to catch some key features of the phenomena and that the results proposed here might provide insights for further works to come.

\section*{Acknowledgements}

The authors would like to thank Alexandra Elbakyan and sci-hub, for removing all barriers in the way of science.

We acknowledge the financial support  by Conselho Nacional de Desenvolvimento Cient\'ifico  e Tecnol\'ogico(CNPq) and Funda\c{c}\~ao Cearense de Apoio ao Desenvolvimento Cient\'ifico e Tecnol\'ogico(FUNCAP) through PRONEM PNE0112-00085.01.00/16.

\end{document}